\documentclass[aps,showpacs,prx,amssymb,amsmath,twocolumn]{revtex4-2}
\usepackage{graphicx}
\usepackage{amssymb}
\usepackage{amsmath}
\usepackage{amsthm}
\usepackage{bm}
\usepackage{xcolor} 
\usepackage{array}
\usepackage{multirow}
\usepackage{tabularx}
\usepackage{booktabs}
\usepackage{textcomp}
\usepackage{mathtools}
\usepackage{hyperref}

\usepackage{qcircuit}

\usepackage{bbm}
\usepackage{cancel}
\usepackage{comment}
\usepackage{physics}

\begin{document}

\title{Quantum tomography of Rydberg atom graphs by configurable ancillas}

\author{Kangheun Kim and Jaewook Ahn}
\address{Department of Physics, KAIST, Daejeon 34141, Republic of Korea}
\date{\today}

\begin{abstract} \noindent
Tomographic reconstruction of the many-body quantum state of a scalable qubit system is of paramount importance in quantum computing technologies. However, conventional approaches which use tomographically orthogonal base measurements require precise and individual qubit controls which are often experimentally daunting. Here, we propose, as a quantum-mechanically robust alternative, to use configurable ancillas of which the continuously-tunable interactions can generate independent base measurements tomographically sufficient for the quantum state reconstruction of the system of interest. Experimental tests are performed for Rydberg atom arrays in $N$-body $W$ states, of which the results demonstrate reliable high-fidelity full quantum state reconstruction of the proposed method. 
\end{abstract}

\maketitle

\section{Introduction} \noindent
Quantum information processing with quantum many-body systems has drawn considerable attention in recent years, because of their promising applications in quantum technologies as well as the fundamental importance of such problems~\cite{NCQC,Sch2012,Noh2016,BrowaeysRev,Daley2022}. As an ultimate tool to evaluate the quantum systems, quantum state tomography (QST) aims to reconstruct the state of a system of interest, using a set of linearly independent and orthogonal measurements on the system~\cite{NCQC,Toninelli2019}. QST is required for all steps of quantum information processing, which include the preparation, manipulation, and measurement steps, respectively requiring the state preparation characterization~\cite{Lvosky2009,Nunn2010,Rambach2021}, quantum process tomography~\cite{Riebe2006,KimYosep2020,Hou2020}, and measurement tomography~\cite{Flu2001,Lundeen2009,Keith2018}. 

Despite the necessity of QST, there are limited experimental demonstrations~\cite{Noguchi2011, Shukla2013, Takeda2021}, especially for a large-scale system. This attributes mainly to the huge size of the tomographic complete set of the orthogonal measurement operators, which grows exponentially as the many-body system size increases. For such a set of measurement operators, QST can use tensor products of Pauli matrices~\cite{NCQC} if qubit operations are near perfect. However, this ideal approach to use the tomographic complete, minimal set of measurements is difficult for a large scale qubit system and the necessary Pauli matrix-based measurement operators require often precise and individual addressing of each qubit~\cite{GrahamSong2022}. 

\begin{figure}[t]
\centering
\includegraphics [width=0.35\textwidth]{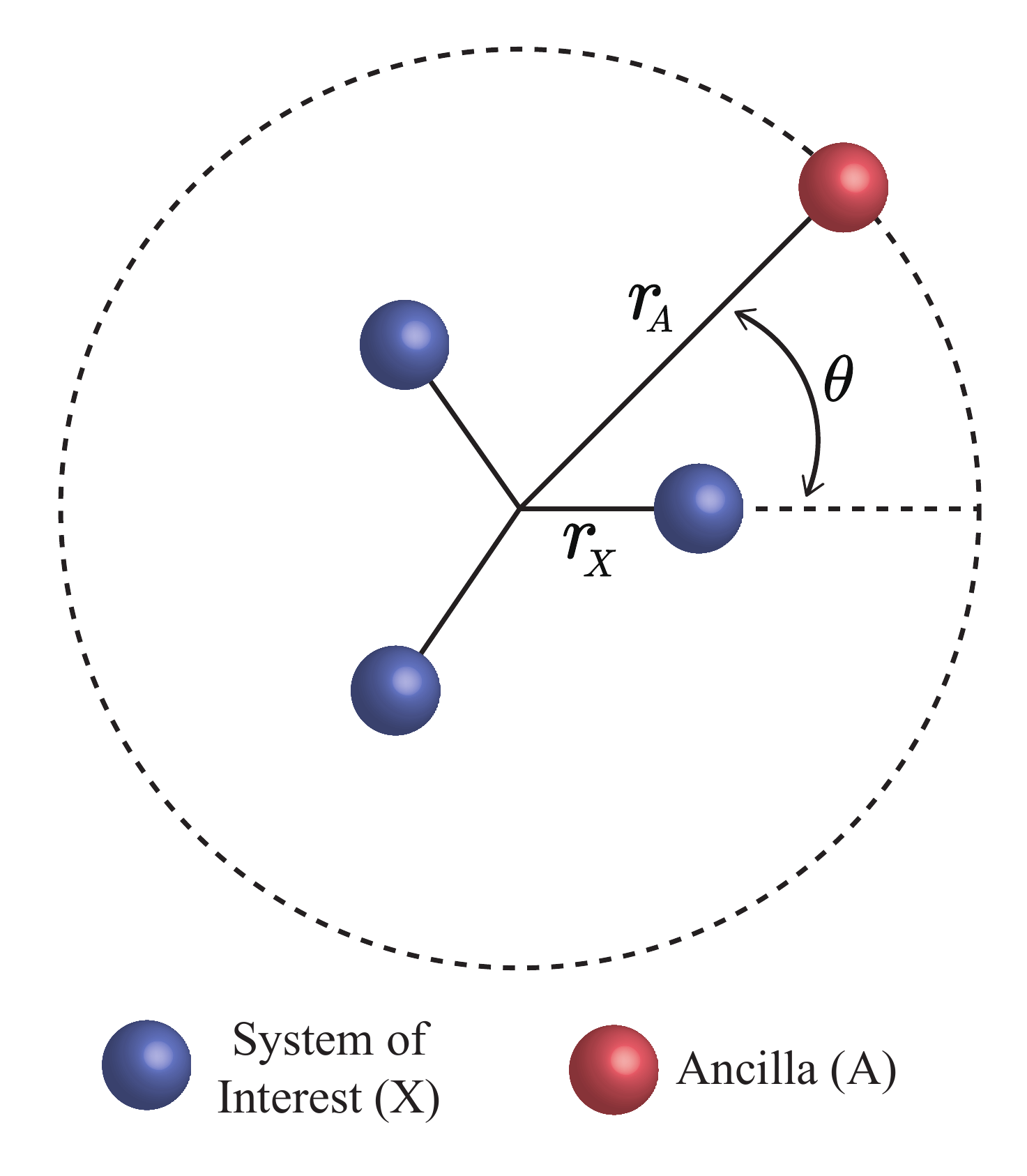}
\caption{A schematic of the quantum state tomography by a configurable ancilla. An ancilla, denoted by $A$, is located on a peripharal location of the system of interest in the center, denoted by $X$, with an adjustable angle $\theta$ and a fixed distance $r$ from the center so that projective measurements of the system and ancilla are performed as a function of $\theta$ for the quantum state tomography of the system.}
\label{Fig1}
\end{figure}

As an alternative, we consider in this paper continuously tunable measurement operators, which are defined with an ancilla (or a system of ancillas) of which the relative location with respect to the system of interest is freely configurable. For example, as shown in Fig.~\ref{Fig1}, one ancillary atom (A) is rotated, bearing resemblance to the computer tomographic scanning, around the set of atoms which are the system of interest (X), so that the relative angular positions of the ancilla generate an independent and continuously tunable set of measurements, with which we perform QST and reconstruct the quantum state of the system. As a prototypical many-body quantum system, we choose to experimentally investigate Rydberg atom systems~\cite{Ebadi2022,Kim2022,Byun2022}, as previous attempts on QST of these systems are done only partially~\cite{Torlai2019} or with the computer simulations~\cite{Zhang2012,Beterov2016,Jo2020}. 

In the rest of this paper, we formulate the working principle of our freely configurable ancilla QST of Rydberg atom arrays in Sec.~\ref{Sec2} and describe the experimental setup and procedure in Sec.~\ref{Sec3}. Experimental results are summarized in Sec.~\ref{Sec4} for $N=2$, 3, 4, and 6 atom systems and compared with numerical most-likely estimation calculations. We then discuss scaling prospects and general applicability of this method to larger and arbitrary Rydberg atom graphs, in Sec.~\ref{Sec5}, before we conclude in Sec.~\ref{Sec6}.

\section{Theoretical Background}
\label{Sec2}
\noindent
We describe the principle, how to use a continuously configurable ancilla for the quantum state tomography of an $N$-body system, i.e., how to generate a sufficient number of independent measurements. As in Fig.~\ref{Fig1}, we consider a system of interest $X$, which is in an unknown state ${\rho}_X$, and an ancilla system $A$. ${\rho}_X$ is a linear combination of orthogonal base states $\{{O}_i\}$, a set of $4^{N}$ matrices satisfying $\text{tr}({O}_i {O}_j)=2^{N} \delta_{ij} $, where $N$ is the number of qubits in $X$.  In the one qubit example, i.e., $N=1$, the orthogonal base states are $\{{O}_i\}=\{\hat{I}, \hat{\sigma}_x, \hat{\sigma}_y, \hat{\sigma}_z \}$. In general, $\hat{\rho}_{X}$ is given by
\begin{equation}
\label{ODCdecomp}
{\rho}_{X} = \sum_{i=1}^{4^{N}} \eta_i {O}_i,
\end{equation}
where the set of the coefficients, $\{\eta_i | i\in \{1,\cdots,4^N\}\}$, determines ${\rho}_X$, so the goal of quantum state tomography is to find  $\{\eta_i\}$.

We introduce a set of measurement superoperators, $\{ \mathcal{M}_k(\rho) | k \in \{1,\cdots,K\} \}$, where $K$ is the number of all measurements and each of the superoperators performs a different measurement on $\rho$. When $\mathcal{M}_k$ is applied on the both hand sides of Eq.~\eqref{ODCdecomp}, we get
\begin{equation}\label{MonODC}
\underbrace{\mathcal{M}_k(\rho_X)}_{P_k} = \sum_{i=1}^{4^{N}} \eta_i \underbrace{\mathcal{M}_k({O}_i)}_{Q_{k}^i},
\end{equation}
where the left hand side is the probability distribution $P_k=\mathcal{M}_k(\rho_X)$, which is obtained experimentally, and the right hand side is the linear combination of quasi-probabilities determined by the operation $\mathcal{M}_k$ acting on the all orthogonal base states. In a linear algebraic form, Eq.~\eqref{MonODC} is given by
\begin{equation}
\begin{pmatrix}
P_{1}\\
\vdots\\
P_{K}
\end{pmatrix}
=
\underbrace{
\begin{pmatrix}
Q_{1}^1&\cdots&Q_{1}^{4^{N}}\\
\vdots&\ddots&\vdots\\
Q_{K}^1&\cdots&Q_{K}^{4^{N}}
\end{pmatrix}}_{\hat{Q}}
\begin{pmatrix}
\eta_1\\
\vdots\\
\eta_{4^{N}}
\end{pmatrix},
\label{Eq3}
\end{equation}
where $\hat{Q}$ is a matrix with $K$ rows and $4^N$ columns. For a non-square matrix, Moore-Penrose inverse~\cite{MPI,Lang2013} can be used to determine the least square estimate solution of the linear algebraic equation; however, the Moore-Penrose inverse gives unique estimate, only for the case where the matrix rank of $\hat{Q}$ is equal to the matrix column number, i.e., $\text{Rank}({\hat{Q}})= 4^{N}$~\cite{LABook}. 
So, more than $4^N$ independent measurements, i.e., $K\ge4^N$ are necessary in Eq.~\eqref{Eq3}.

Measurement superoperators $\mathcal{M}$ can be defined for Rydberg atom systems of which the many-body Hamiltonian is an Ising spin Hamiltonian, where the peudo-spin states, $\ket{0}$ and $\ket{1}$, represent the ground and Rydberg atoms, respectively. In the unit of $\hbar=1$, the Hamiltonian for the $X$ and $A$ atoms is given by
\begin{equation}
\label{HRyd}
\hat{H} = \frac{\Omega}{2}\sum_{i\in X\cup  A}{\hat{\sigma}_x^{(i)}} - \frac{\Delta}{2}\sum_{i\in X\cup A}{\hat{\sigma}_z^{(i)}} +\sum_{i<j}{V_{ij}\hat{n}_i \hat{n}_j},
\end{equation}
where $\Omega$ is the Rabi frequency, $\Delta$ is the detuning, $V_{ij}=C_6/{r}_{ij}^6$ is the van der Waals interaction between two Rydberg atoms, and $\hat{n}=(\hat{\sigma}_z+1)/2$ is the Rydberg excitation. With the Hamiltonian, the superoperators are defined as
\begin{equation}\label{Mdef}
\mathcal{M}_{n,\theta}(\rho_{X}) = \text{tr}\left(\hat{\Pi}_n\hat{U}(\theta)\left(\rho_X\otimes\ket{0}_A\bra{0}_A\right)\hat{U}^{\dagger}(\theta)\hat{\Pi}_n\right),
\end{equation}
where $n$ is the bitstring for the spin configuration of the atoms in $X\cup A$, $\theta$ is the angular position of the ancilla, $\Pi_n$ is the projective measurement onto the $n^\text{th}$ bitstring spin basis, $\ket{0}_A$ is the initial spin state of the ancilla, $U(\theta)$ is the time evolution of $\rho_{X\cup A}$ by $\hat{H}$ that depends on $\theta$. As $\theta$ can be continuously adjustable, the angle-dependent time evolution $\hat{U}(\theta)=\exp(-i\hat{H} (\theta) t_{E})$, for a sufficient time $t_E$, results in an entanglement of $X$ and $A$ differently for a different angle $\theta$~\cite{Nieu2004}, so we can generate sufficiently many superoperators, i.e., $K=2^{N+1} \times \norm{\{\theta\}}\ge4^N$, which are in general independent with each other.

\begin{figure*}[bt]
\centerline{\includegraphics [width=0.95\textwidth]{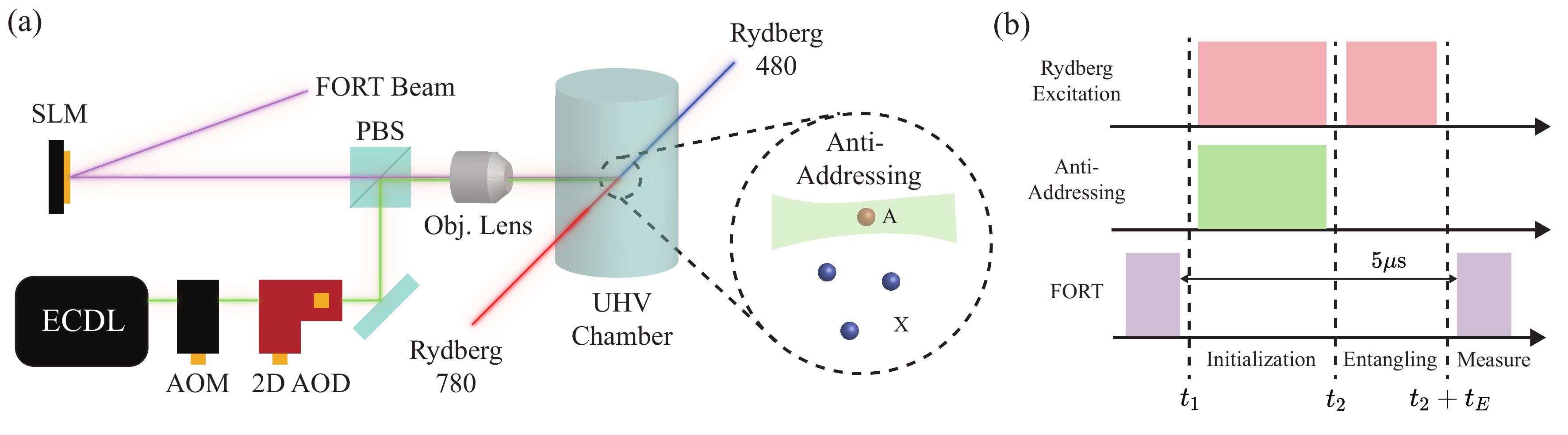}}
\caption{Rydberg-atom programmable quantum simulator. (a) Cold rubidium atoms are trapped with a far-off resonant optical traps (FORT), which are holographically controlled by a spatial light modulator (SLM), and optically excited to the Rydberg state by counter-propagating two-photon transition beams, which are suppressed by the anti-addressing beam controlled by an acousto-optic modulator (AOM) and a two-dimensional acousto-optic deflector (AOD). (b) Experimental pulse sequence.}
\label{Fig2}
\end{figure*}

For a limiting case of weakly interacting ancilla, we can consider the independence of angular measurements. When we place the ancilla far away from the system, i.e., $r_A\gg r_X$, the Hamiltonian in \eqref{HRyd} can be written with three parts, 
\begin{equation}
\hat{H} = \hat{H}_X + (\Omega/2)\sigma_x^{(A)}+ \sum_{x\in X}V_{xA}\hat{n}_x\hat{n}_A,
\end{equation}
where the first $\hat{H}_X$ is the Hamiltonian of $X$, and the second is the Rabi oscillation of the ancilla, and the last is the interaction between $A$ and $X$. Because $V_{xA}\propto (r_{A}-r_x)^{-6}\ll 1$, the first order perturbation of $\hat{U}(\theta)$ is given by
\begin{equation}
\label{Ufar}
\hat{U}(\theta) \approx e^{-i(\hat{H}_X + (\Omega/2)\sigma_x^{(A)})t_{E}} +i \sum_{x\in X}{V_{xA}(\theta)\hat{\mathcal{I}}}_x
\end{equation}
where $\hat{\mathcal{I}}_x$ is a geometry independent integral term dependent only on $t_E$. When Eq.~\eqref{Ufar} is used in Eq.~\eqref{Mdef}, the coefficient $Q_{n,\theta}^i$ of the matrix $\hat{Q}$ is given by
\begin{equation}
Q_{n,\theta}^i\approx c_i +\sum_{x\in X}{V_{xa}(\theta)v_{x,i}},
\end{equation}
where $c_i$ and $v_{x,i}$ are constants for each orthogonal matrix $O_i$. As $V_{xa}(\theta)$ changes for each $\theta$, the two rows of the $\hat{Q}$ matrix differs. For example, with different $\theta_1$ and $\theta_2$, and with the same projection $\hat{\Pi}_n$, the two row vectors of the matrix $\hat{Q}$ and their difference is given by
\begin{equation}
\vec{Q}_{n,\theta_2} - \vec{Q}_{n,\theta_1} =
\sum_{x\in X}{\left\{V_{xA}(\theta_2) - V_{xA}(\theta_1)\right\}} \vec{v}_x,
\end{equation}
where $\vec{v}_x= \begin{bmatrix}v_{x,1} && v_{x,2} && \cdots && v_{x,4^N}\end{bmatrix}$ is a constant vector for each $x$; thus the two row vectors $\vec{Q}_{n,\theta_2}$ and $\vec{Q}_{n,\theta_1}$ are independent for $\theta_1\neq \theta_2$. Therefore the rank of the matrix $\rm{Rank}(\hat{Q})$ can be increased as much as we want by measuring with different angle $\theta$s, until the maximum value $\rm{Rank}(\hat{Q})/4^N=1$.

\section{Experimental procedure} \label{Sec3}
\noindent
Experimental quantum state tomography of Rydberg atom systems is conducted with a Rydberg-atom programmable quantum simulator. A schematic of the experimental setup is shown in Fig.~\ref{Fig2}(a), which is primarily similar to the one previously reported elsewhere~\cite{Woojun2019,Kim2022,Byun2022}, with an additional setup assembled to locally address individual atoms~\cite{Labuhn2014}. 
We utilized cold rubidium atoms ($^{87}\text{Rb}$) arranged in the two- or three dimensional space with a set of optical tweezers (i.e., far off-resonant optical dipole traps, FORT). The atoms were initially cooled in a magneto optical trap (MOT) down to a temperature of $\sim$30~$\mu$K, via Doppler and polarization gradient coolings (PGC) and loaded into the optical tweezers individually. Both the qubit atoms and the ancilla atom (or atoms) were deterministically prepared to wanted target places by the atom rearrangement procedure~\cite{Hyosub2016,Hyosub2018}, and optically pumped to the ground state $\ket{0}=\ket{5S_{1/2},F=2,m_F=2}$. Then, the atoms were excited to the Rydberg state, $\ket{1}= \ket{71S_{1/2},J=1/2,m_J=1/2}$, by a two-photon laser excitation, $\ket{0} \rightarrow  \ket{5P_{3/2},F=3,m_F=3} \rightarrow \ket{1}$.

Qubit operations were performed by the Rydberg excitation of the system atoms, in such a way that individually-addressed atoms are Rydberg-excitation suppressed, i.e., anti-addressed, as shown in Fig.\ref{Fig2}(a). The anit-addressing beam was spatially modulated with a 2D acousto-optic deflector (2D-AOD, AA Opto Electronics DTSXY-400-800) and switched on and off at a time scale of a few nanosecond rise and fall times with an acousto-optic modulator (AOM). The 2D-AOD split and deflects the beam in a wide range for a set of anti-addressing beams, which were used to off-detune the Rydberg-atom excitation of chosen atoms, i.e., applying very large $\Delta$ via the AC Stark shift. The as-prepared anti-addressing beams were then merged with the optical tweezers which were sculpted by a spatial light modulator (SLM, ODPDM512, Meadowlark Optics) before being sent to the atoms in the ultra-high vacuum chamber. The experimental pulse sequence for laser operations is shown in Fig.~\ref{Fig2}(b). The optical tweezer beams were first all turned off at $t_1$ and the Rydberg excitation of both the system atoms and ancilla is tuned on globally, while the anti-addressing beam for the ancilla was turned on locally, so that the total system $X\cup A$ is prepared as ${\rho}_{X\cup A}(t_2)={\rho}_{X}\otimes \ket{0}\bra{0}_{A}$. Then with the anti-addressing off, Rydberg excitation is again performed globally to entangle the total system, i.e., ${\rho}_{X\cup A}(t=t_2+t_E)=\hat{U}(\theta){\rho}_{X\cup A}(t_2) \hat{U}^\dagger(\theta)$. 

After the qubit operations, qubit measurements were carried out by taking an image of remaining atoms in $\ket{0}$ with an electron multiplying charge-coupled device (EMCCD), after other atoms in $\ket{1}$ were anti-trapped by the optical tweezers. The entire steps of the above process were repeated with various relative angle $\theta$ between the system and the ancilla to collect the projection measurements $\Pi_n$ of the total system, i.e., obtaining $P_{n}(\theta) = \mathcal{M}_{n,\theta}({\rho}_X)$ probability values, which were then used to reconstruct the tomographic information ${\rho}_{X}$ of the system.

\section{Experimental Results}
\label{Sec4} 
\begin{figure}[t]
\centering
\includegraphics [width=0.48\textwidth]{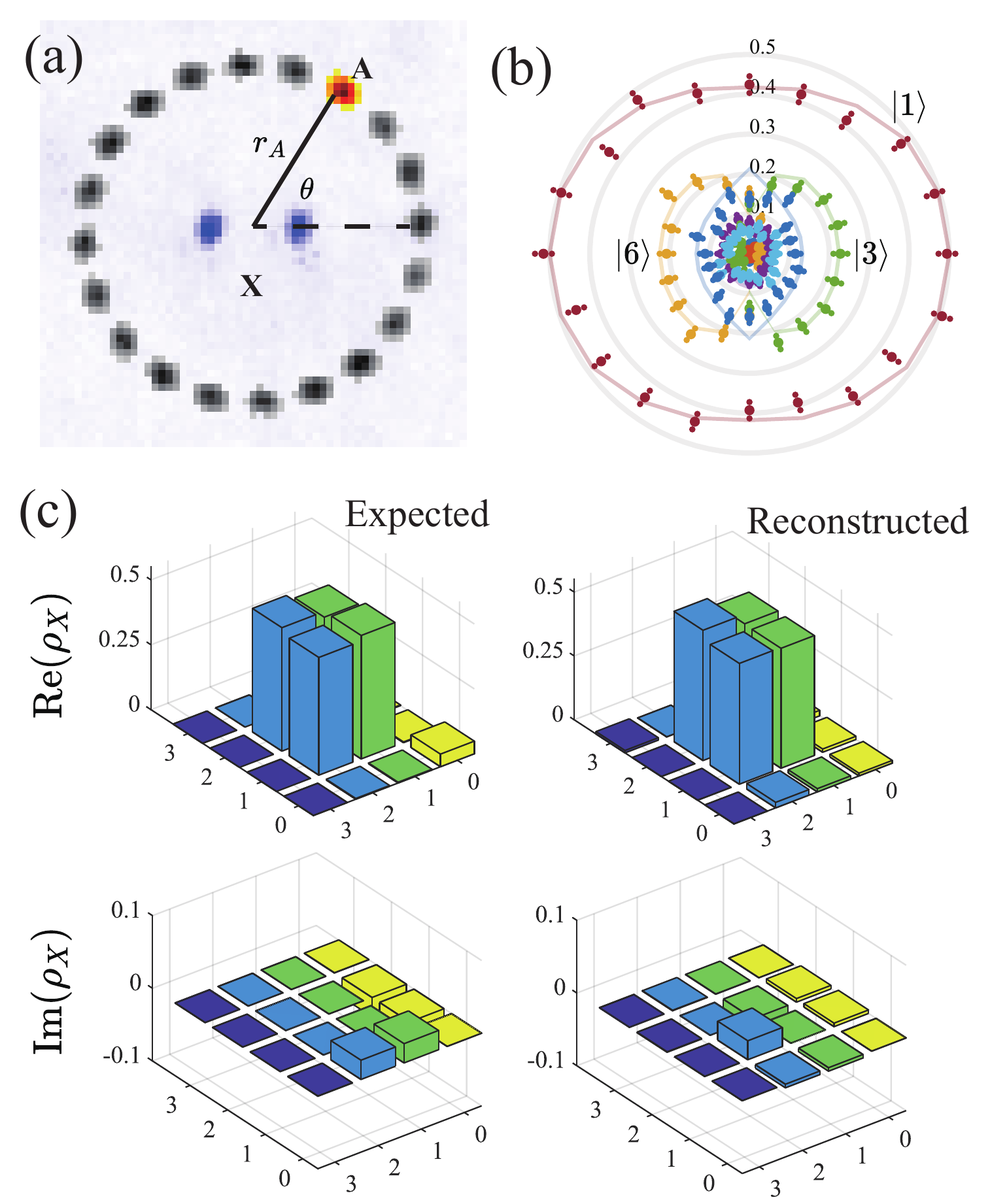}
\caption{Quantum state tomography of two atoms in Bell state. (a) Atom images of the system (in blue) and ancilla (in red or gray). (b) Measured probabilities, $P_{n}(\theta)=\mathcal{M}_{n,\theta}({\rho}_X)$, in the polar coordinate system, for chosen three-spin bitstring ($n$) configurations, where the red data are for $\ket{n=1}=\ket{00}_X\otimes \ket{1}_A$; yellow for $\ket{n=5}=\ket{10}_X\otimes \ket{1}_A$; green for $\ket{n=3}=\ket{01}_X\otimes \ket{1}_A$; and blue for $\ket{n=0}=\ket{00}_X\otimes \ket{0}_A$. (c) Comparison between the theoretically expected density matrix $\rho_X^{\rm Cal}$ (left) and experimentally reconstructed $\rho_X^{\rm QST}$ (right), with $x,y$ coordinates representing the bitstring $n$.
}
\label{Fig3}
\end{figure}

\begin{figure*}[t]
\centerline{\includegraphics [width=1\textwidth]{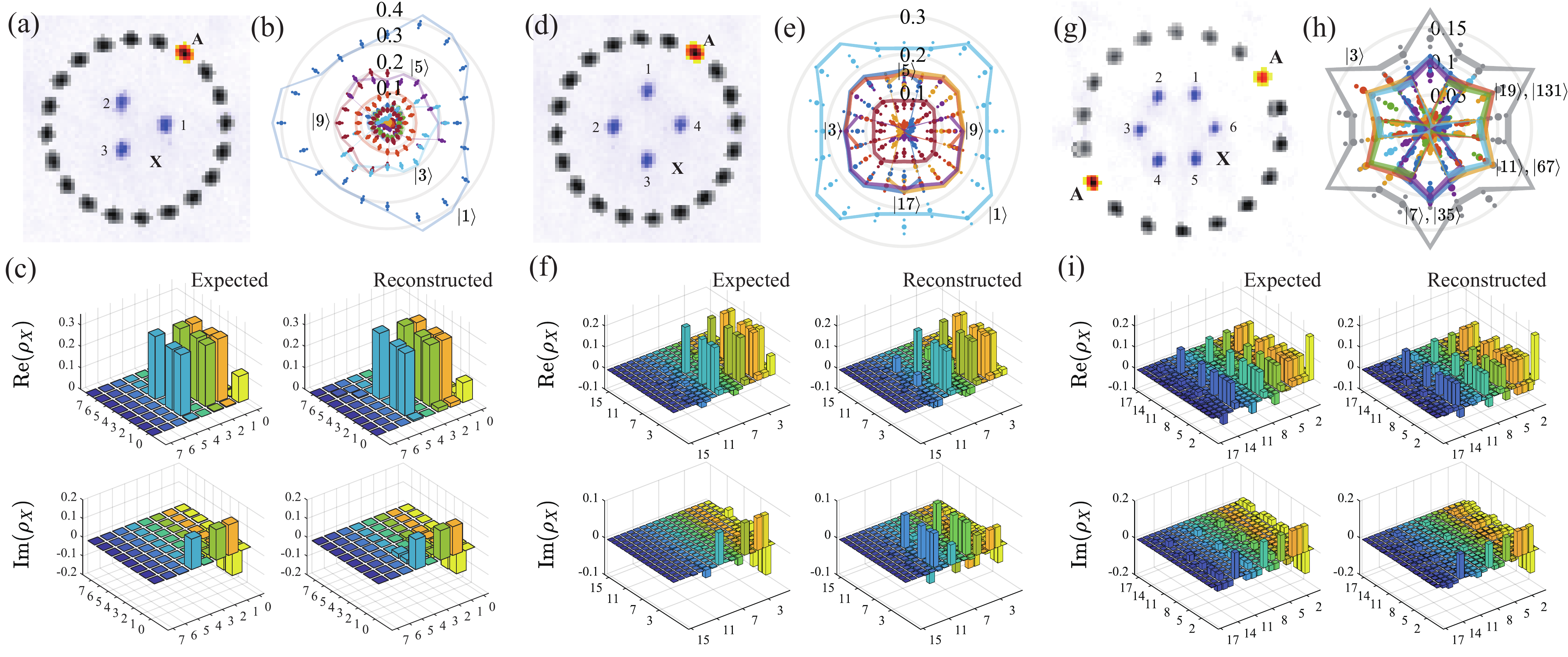}}
\caption{Quantum state tomography of (a-c) three, (d-f) four, and (g-i) six atoms in their $W$ states. (a,d,g) Post processed atom images for (a) $N=3$ and $N_A=1$, (d) $N=4$ and $N_A=1$, and (g) $N=6$ and $N_A=2$. (b,e,h) Experimented probability data with dots representing averages and small dots error bars, where the lines are theoretically expected probabilities and different colors denote the probabilities of different bitstring states. (b) $N=3$ probabilities, $P_{{1}}(\theta)$ (dark blue), $P_{{9}}(\theta)$ (red), $P_{{5}}(\theta)$ (purple), $P_{{3}}(\theta)$ (sky blue), and $P_{{0}}(\theta)$ (orange). (e) $N=4$ probabilities, $P_{1}(\theta)$ (light blue), $P_{17}(\theta)$ (purple), $P_{9}(\theta)$ (yellow), $P_{5}(\theta)$ (orange), $P_{3}(\theta)$ (dark blue), and $P_{0}(\theta)$ (red). (h) $N=6$ probabilities, $P_{3}(\theta)$ (grey), $P_{131}(\theta)$ (orange), $P_{19}(\theta)$ (green), $P_{35}(\theta)$ (blue), $P_{7}(\theta)$ (purple), $P_{67}(\theta)$ (yellow), and $P_{11}(\theta)$ (sky blue). (d,f,i) Theoretically expected density matrix $\rho_X^{\rm Cal}$ (left) and reconstructed density matrix $\rho_X^{\rm QST}$ (right) with $x,y$ coordinates representing the bitstring $n$ of each experiment.}
\label{Fig4}
\end{figure*}

\noindent
Four different systems ($X$) of $N=2$, 3, 4, and 6 atoms are considered for experimental quantum state tomography, of which the results are summarized in Fig.~\ref{Fig3} for the $N=2$ experiment and in Fig.~\ref{Fig4} for the $N=3$, 4, and 6 experiments. For the first three experiments ($N=2,3,4$), we used one ancilla ($N_A=1$) and for the last $N=6$ experiment, we used two ancillas ($N_A=2$).

In the first experiment ($N=2$), the system of interest, $X$, is a two-atom system, as shown in the center (in blue color) of Fig.~\ref{Fig3}(a), and one ancilla atom $A$ is used, for example, as shown in red in the same figure. The other angular positions of the ancilla atom are shown in gray for reference. We performed 20 angular measurements with $\theta_j =2\pi/k$ for $j=1,\cdots, 20$, and kept the distance of the ancilla the same from the center of $X$, to satisfy $K=20 \times 2^{N+1} > 4^N$. We chose as a known state ${\rho}_X$ for the quantum state tomography, the Bell state, i.e.,
\begin{equation}
{\rho}_{X} \approx \ket{\psi_{\rm Bell}} \bra{\psi_{\rm Bell}},
\end{equation} 
where $\ket{\psi_{\rm Bell}}=(\ket{01}+\ket{10})/\sqrt{2}$, by placing the two atoms of $X$ within the Rydberg blockade distance,  $r_b=(\hbar C_6/\Omega)^{1/6}\approx 10~\mu\text{m}$. 

The polar plot in Fig.~\ref{Fig3}(b) shows measured probabilities, $P_{n}(\theta)=\mathcal{M}_{n,\theta}({\rho}_X)$, as a function of $\theta$, where each color depicts a different bitstring three-spin state $n$. Observed probabilities are significant for states $\ket{n=1}=\ket{00}_X\otimes \ket{1}_A$, $\ket{3}= \ket{01}_X\otimes \ket{1}_A$, $\ket{6}=\ket{10}_X\otimes \ket{1}_A$, and $\ket{0}= \ket{00}_X\otimes \ket{0}_A $. Their probabilities $P_{1}(\theta)$ (red), $P_{3}(\theta)$ (yellow), $P_{6}(\theta)$ (green), $P_{0}(\theta)$ (blue) are shown in Fig.~\ref{Fig3}(b) along with others. The angular behavior of the probabilities can be understood with the Rydberg blockade effect; for example, $P_{3}(\theta)$ (yellow) is significant in the angular region $\{\pi/2<\theta<3\pi/2\}$, where the second atom in $X$ is not Rydberg-blockaded by the ancilla, and $P_{6}(\theta)$ (green) is almost zero in the same region because the first atom is Rydberg-blockaded, etc. About 550 experiments were repeated for each angle measurement. Other experimental parameters are summarized in Table~\ref{table:tb1}, which include the system and ancilla atom distances from the center, Rabi frequencies, and the pulse lengths, the initialization and entanglement times, $t_{ I}=t_2-t_1$ and $t_E$, respectively. 

\begin{table}[b]
\caption{Experimental parameters. $N$ and $N_A$ are the atom numbers in $X$ and $A$, $r_X$ and $r_A$ are the distances of $X$ and $A$ atoms in Figs.~\ref{Fig3} and \ref{Fig4} from their center, $\Omega$ is the Rabi frequency, and $t_{I}=t_2-t_1$ and $t_{E}$ are pulse lengths in Fig.~\ref{Fig2}(b) for initialization and entanglement, respectively.}
\label{table:tb1}
\centering
\begin{ruledtabular}
\begin{tabular}{ccccccc}
$N$ & $N_A$ & $r_X$ ($\mu$m) & $r_A$ ($\mu$m) & $\Omega$ (MHz)  & $t_{I} $ ($\mu$s) & $t_{ E}$ ($\mu$s)\\
\hline
$2$ & 1 &2.5& 9 &0.896  &0.387 &0.595 \\
$3$ &1 & $5/\sqrt{3}$ &9 &0.894  &0.259 &0.595 \\
$4$  &1 &4 &10 &0.855 &0.275 &0.563 \\
$6$  &1 & 4.5 &12 &0.845 &0.211 &0.579 \\
\end{tabular}
\end{ruledtabular}
\end{table}

Figure~\ref{Fig3}(c) shows the resulting $\rho_X^{\rm QST}$ (right) of the experimental quantum state tomography in comparison with a theoretically calculated $\rho_X^{\rm Cal}$ (left). The result shows a high fidelity
\begin{equation}
\mathcal{F}(N=2)=\text{tr}\sqrt{\sqrt{{\rho}_{X}^{\rm Cal}}{\rho}_{X}^{\rm QST}\sqrt{{\rho}_{X}^{\rm Cal}}}=0.976(9)
\end{equation}
of the quantum state tomography of the two atoms in Bell state. For the theoretically calculated ${\rho}_X^{\rm Cal}$ (left), experimental errors are taken into account which include the state preparation and measurement errors, individual and collective dephasing contributions, and control errors (see Appendix~\ref{appendix:error} for details). For the experimental QST $\rho_X^{\rm QST}$, we used Bayesian mean estimation (BME) with Markov chain Monte Carlo (MCMC) method~\cite{Blume2010,Struchalin2016} (See Appendix~\ref{appendix:BME} for details), in order to avoid nonphysical $\rho_X^{\rm QST}$ (i.e., of negative probabilities) which is unavoidable due to experimental and statistical errors~\cite{Granade2017,Lang2013}. 

Similarly, $N=3$, 4, and 6 atom experiments are shown in Fig.~\ref{Fig4}. The unknown states were chosen respectively near their $N$-qubit $W$-states, i.e.,
\begin{subequations}
\begin{eqnarray}
\ket{\psi}_X^{N=3}&\approx&\frac{1}{\sqrt{3}} (\ket{001}+\ket{010}+\ket{100}), \\
\ket{\psi}_X^{N=4}&\approx&\frac{1}{2} (\ket{0001}+\ket{0010}+\ket{0100}+\ket{1000}), \\
\ket{\psi}_X^{N=6}&\approx&\frac{1}{\sqrt{6}} (\ket{000001}+\cdots+\ket{100000}).
\end{eqnarray}
\end{subequations}
for the $N=3$, 4, and 6 atom experiments, receptively. $N=3$, 4 experiments are summarized in Figs.~\ref{Fig4}(a-c) and Figs.~\ref{Fig4}(d-f), respectively, where one ancilla atom is used as shown in Figs.~\ref{Fig4}(a,c), while for the $N=6$ atom experiment in Figs.~\ref{Fig4}(f-h) two ancilla atoms are used as shown in Fig.~\ref{Fig4}(f). Experimentally measured probabilities, $P_{n}(\theta)$, are shown in Figs.~\ref{Fig4}(b,d,g) for $\rho_X^{N=3}$, $\rho_X^{N=4}$, and $\rho_X^{N=6}$, respectively. The 120$^\circ$ (for $N=3$), 90$^\circ$ ($N=4$), and 60$^\circ$ ($N=6$) dependence of angular probability measurements are clearly shown, as expected from their Rydberg blockade nature between constituent atoms in $X$ and $A$. About 1300, 550, and 700$\sim$1000 experiments were repeated for each angle measurement of  $N=3$, 4, and 6, respectively. In all cases, the results of the quantum state tomography in Figs.~\ref{Fig4}(c,e,h) show high fidelities $\mathcal{F}(N=3)=0.975(3)$, $\mathcal{F}(N=4)=0.88(1)$, and $\mathcal{F}(N=6)=0.85(1)$ of the quantum state reconstructions. We note that, for the sake of calculation convenience of the $N=6$ experiment which is performed in a strong Rydberg blockade condition, i.e., $r_X = 4.5~\mu\text{m} \ll r_b \approx 10~\mu\text{m}$, we ignored anti-blockaded states of nonzero adjacent double excitations.

\section{Discussion}
\label{Sec5} 
\begin{figure}[t]
\centering
\includegraphics [width=0.45\textwidth]{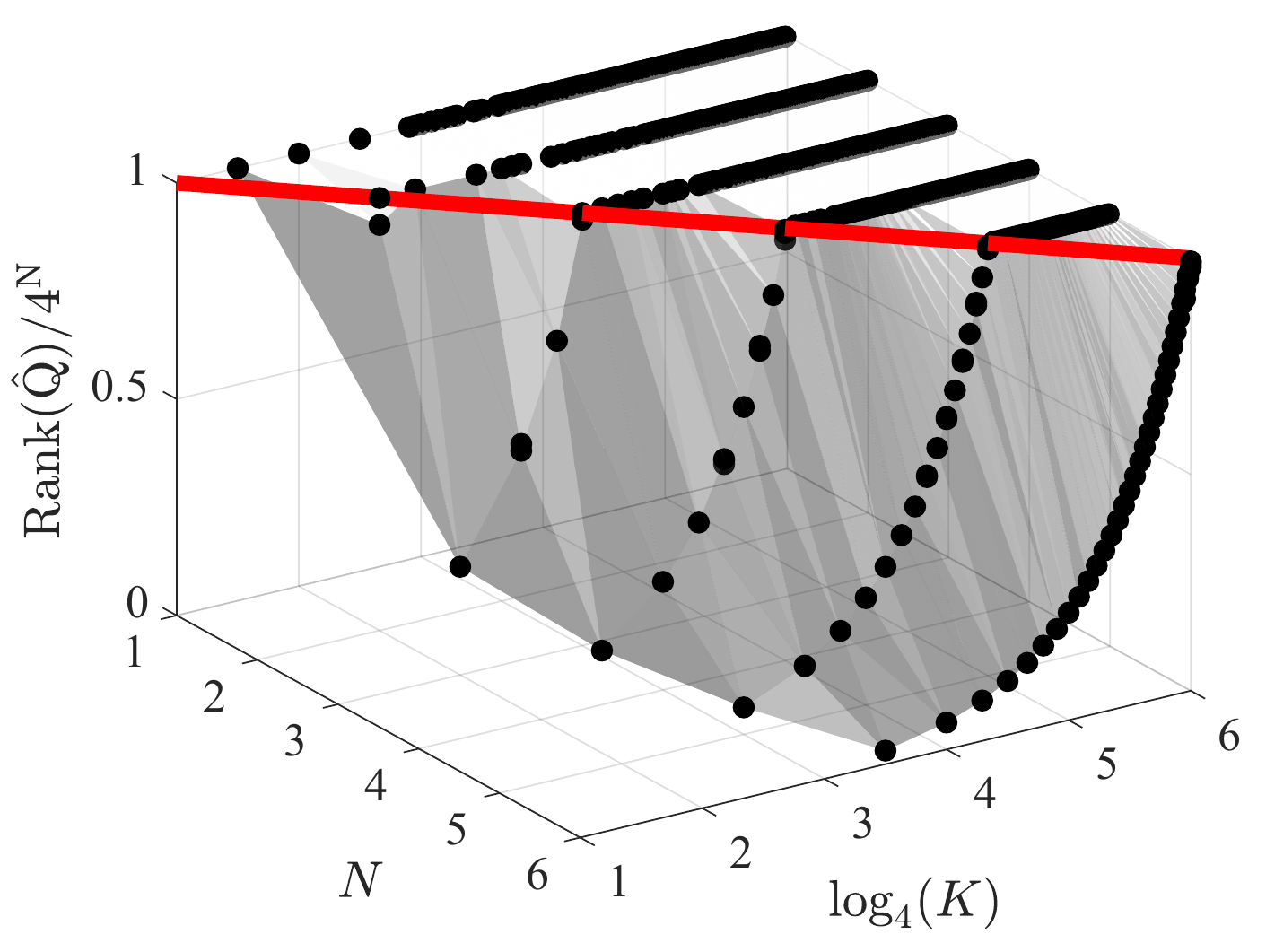}
\caption{$\rm{Rank}(\hat{Q})/4^N$ value with $N$ and $K$, simulated with 5000 random geometries of $X$ and $A$. The minimum measurement condition, $K=4^N$, is drawn with the red line. The random graphs of which the measurement number exceeds the minimum condition, $K\ge 4^N$, satisfy the sufficient condition for full QST reconstruction, i.e., $\rm{Rank}(\hat{Q})/4^N=1$, which implies the independence of the arbitrary configurable ancilla measurement.}
\label{Fig5}
\end{figure}

\noindent
While our experimental tests of the continuously-configurable ancilla are successfully performed for a set of strongly-symmetric system structures, it is worthwhile to consider general graphs of random system geometries as well as random ancilla positions (of different distances from the system as well). As in Sec.~\ref{Sec2}, $\rm{Rank}(\hat{Q})/4^N=1$ is a good measure of measurement ($\mathcal{M}$) independence, which is necessarily satisfied by a sufficient $K>4^N$, so we numerically test if the condition $K>4^N$ is sufficient as well. We first parameterize the position of ancilla system $A$ with an arbitrary geometry $\{G\}$ instead of $\{\theta\}$, as the ancilla position is not necessarily bounded to the circle as in Fig.~\ref{Fig1}. Random graph geometries of system $X$ and ancilla $A$ with the fixed state $\ket{0\cdots 0}_A$ were generated by the algorithm 2 of Ref.\cite{Serret2020} with density $\nu = 0.1$ and exclusion radius $r_{\rm exc}=5~\mu\text{m}$. The parameters for the entanglement of the two systems $X$ and $A$ were $\Omega = 1$ MHz and $t_{E}=t_{\pi/2}$. After each random graph was generated, we calculated $\hat{Q}$ matrix for each geometric case and computed $\text{Rank}(\hat{Q})$ numerically. In Fig.~\ref{Fig5}, simulated results are shown for 5000 random graphs, where $x$-axis is $N$, $y$-axis is $\log_4{K}$, and $z$-axis is the $\text{Rank}(\hat{Q})/4^N$. The red line in Fig.~\ref{Fig5} is the $K = 4^N$ condition. So, it is clear that at least for $N\le 6$, the method using a freely-configurable ancilla generates enough independent measurements, and, as a result, $K \ge 4^N$ is the sufficient condition for $\text{Rank}(\hat{Q})/ 4^N=1$. 

We now discuss the scalability of our method for a large-scale graph of atoms. Under the assumption that the independence of measurements $\rm{Rank}(\hat{Q})/4^N=1$ is satisfied for a large $N$, possible technical issues are (1) experimental time and (2) (classical) computational power. First, as for the experimental time, the minimally required different geometric positions are $\norm{\{G\}}_{\rm min}=2^{N-N_A}$, where we consider general geometries $\{G\}$ instead of $\{\theta\}$ for a general $K$, and $N_A$ the ancilla atom number. As in Refs.~\cite{Nieu2004,Shukla2013}, increasing the ancilla number $N_A$ will reduce the required geometry number. However for each angular point, to maintain the ratio of standard deviation and the probability of the multinomial distribution of the experiment, the number of requiring experimental repetition scales as $\sim 2^{N+N_A}$ for each angle; therefore, the total required experimental number of repetition scales $\sim 2^{N-N_A}\times 2^{N+N_A}=4^N$, not dependent on the ancilla number. As a result, using one ancilla, i.e., $N_A=1$, is essentially the same as increasing $N_A$, so our method is valid with one ancilla. Second, our tested computational time of the numerical reconstruction simply exceeds computation limit of desktop computing machines, as a $4^N$ parameter fitting is required which exponentially increases with the system atom number $N$. A desktop memory of 128 GB is not enough for $N=18$ and a supercomputer memory (Fugaku, RIKEN, 32 GB/node $\times$ 158,976 nodes) for $N=26$. As for the computational time, e.g., a BME time complexity of an advanced algorithm~\cite{Lukens2020} scales as $\sim O(4^N)$, which makes an $N=12$ BME difficult, taking a week of computing estimation, for the desktop (2.5 GHz single thread) and an $N=25$ for a supercomputer (Fugaku, RIKEN, 442 PFLOPS, assuming $\sim5$ GFLOPS/core for desktop). Parallel computing with an advanced GPU algorithm~\cite{GPUtomo}, reducing the fitting parameters by matrix product states (MPS)~\cite{Kurmapu2022}, or neural networks~\cite{Torlai2019} can be considered for this computational time and memory issues.

\section{Conclusion} \label{Sec6} \noindent
In summary, we have performed a full state tomography of the Rydberg atom systems, using a freely arrangeable ancillary atom system. The use of continuously tunable angular position of the ancilla is an unique advantage of an optical tweezer system, allowing easy generation of a sufficient number of independent measurements that were tomographically enough for the quantum state reconstruction. In experiments, we
tested 2, 3, 4, 6 atom systems either in Bell or $W$ state. With a numerical analysis of random graphs, we show that the quantum state tomography is robust against the target and ancilla atom positions. It is estimated that the presented QST method is applicable for up to $N\approx 26$, mainly limited by the memory capacity and time complexity of state-construction classical computation.

\begin{acknowledgements} \noindent
This research was supported by Samsung Science and Technology Foundation (SSTF-BA1301-52) and National Research Foundation of Korea (2017R1E1A1A01074307).
\end{acknowledgements}

\appendix

\section{Experimental error calibration}
\label{appendix:error}
\noindent
Experimental systematic errors which include global and individual dephasing errors and the state preparation and measurement errors (SPAM) are calibrated with the procedure previously discussed in Refs.~\cite{Woojun2019,WoojunDoc}. The individual dephasing of the system and ancilla  atoms attributes to the spontaneous decay from the intermediate state $\ket{5P_{3/2}, F=3, m_F=3}$ to $\ket{0}$. A phenomenological Lindbaldian equation
is used to numerically estimate the corresponding dephasing coefficient $\gamma_{\rm ind}=20 \text{kHz}$ for all experiments. The collective dephasing of coefficient $\gamma_{\rm col}$ is also numerically estimated using the phase noise spectrum of the laser systems. State praparation errors are negligibe, as the optical pumping fidelity is measured to be over 98\%. Measurement errors are numerically estimated with $\chi^2$ fittings of experimental measurements~\cite{Tuch2008,ErrorBook} and the as-obtained bit-flip conditional probabilities, $P(0|1)$ and $P(1|0)$ from $\ket{1}$ to $\ket{0}$ and from $\ket{0}$ to $\ket{1}$, respectively, are listed in Table~\ref{table:tb2}  along with other fitted errors. The simulated results in Sec.~\ref{Sec3} were considering the fixed $\gamma_{ind}$ and fitted $\gamma_{col}$, and the experimental results in Sec.~\ref{Sec3} are SPAM-corrected~\cite{Les2018} with the matrix $P$ given by
\begin{equation}
\bigotimes_{i=1}^N
\begin{pmatrix}
1-P(0|1)+P(1|0)P(0|1)& P(1|0)\\
P(0|1)-P(1|0)P(0|1) & 1-P(1|0)
\end{pmatrix}.
\end{equation}

\begin{table}[b]
\caption{Experimental Rabi frequency, collective dephasing coefficient, and the state preparation and measurement errors}
\begin{ruledtabular}
\centering
\begin{tabular}{cllll}
$N$ & $\Omega$ (MHz) & $\gamma_{\rm col}$ (MHz) & $P(1|0)$ & $P(0|1)$\\
\hline
2 &$0.896(9)$ & $0.07(3)$ & $0.02(4)$ & $0.10(4)$\\
3 &$0.894(8)$ & $0.05(3)$& $0.03(3)$ & $0.12(3)$  \\
4 &$0.855(11)$ &$0.11(3)$ &$0.01(4)$ &$0.10(4)$  \\
6 &$0.845(7)$ &$0.05(2)$ &$0.03(3)$ & $0.12(3)$ \\
\end{tabular}
\end{ruledtabular}
\label{table:tb2}
\end{table}

\section{Bayesian mean estimation details}
\label{appendix:BME} 
\noindent
The method of Bayesian mean estimation (BME) is used to obtain the solution parameters $\{\eta_i \}$ in Eq.~\eqref{Eq3} which most-likely explain the experimental results which is widely used in estimation of parameter using quantum experimental data including QST, Hamiltonian learning, etc.~\cite{Hus2012,Krav2013,Struchalin2016,Wang2017}. The following explanation and the algorithm mostly follows Ref.~\cite{Blume2010}. The likelihood function is defined with the experimental data ($E_{m,n}$) and the model ($\rho$) as
\begin{equation}
\mathcal{L}(\rho) = P_{1,1}(\rho)^{E_{1,1}}P_{1,2}(\rho)^{E_{1,2}}\cdots P_{M,2^{N}}(\rho)^{E_{M,2^N}},
\label{EqL}
\end{equation}
where each $P_{m,n}(\rho)$ denotes the model probability and $E_{m,n}$ is the number of measurement of the $n^{\rm th}$ state in the $m^{\rm th}$ arrangement of the ancilla atom. As the simple maximization of $\mathcal{L}(\rho)$, i.e., the maximal likelihood estimation (MLE), have some problems; (1) it often results in the non-realistic $\rho$ estimation with the $0$ eigenvalues (2) it does not have a natural error on the estimation. We instead use BME with constraints of a physical density matrix, i.e., for each density matrix in the density matrix domain, we consider the probability to be a real solution density matrix. 

BME assumes a prior probability distribution $\pi_0(\rho)$ in such a way that a posterior probability distribution, with experimental results taken into account, is given by $\mathcal{L}(\rho)\pi_0(\rho)$. Then, the solution density matrix, $\rho_{\rm BME}$ is estimated by integrating the posterior probability over the whole density matrix space, i.e.,
\begin{equation}
{\rho}_{\rm BME} = \int \rho\mathcal{L}(\rho)\pi_0(\rho) d\rho.
\end{equation}
In our calculations in Sec.~\ref{Sec3}, we sampled, for the prior $\pi_0(\rho)$, a pure state $\ket{\psi}$ from the $2^{N+1}$ dimensional Hilbert space with Haar measure and traced out the auxiliary dimension to get $\rho = \text{tr}_a(\ket{\psi}\bra{\psi})$. As it is daunting to calculate the integral for the entire density matrix space, we used the Markov chain Monte Carlo (MCMC) method, which uses a jumping algorithm to sample multiple $\rho$s according to the posterior probability distribution $\mathcal{L}(\rho)\pi_0(\rho)$. We adopted the jumping algorithm in, the Metropolis-Hastings (MH) algorithm, with a jumping step size optimized for fast fidelity convergence. Jumping step size was chosen $\Delta = 4\pi \Delta_0 \langle T(\rho,\langle \rho \rangle) \rangle$ similar to Ref.~\cite{Struchalin2016} where constant $\Delta_0=0.1$ was chosen and the $\langle T(\rho,\langle \rho \rangle)\rangle$ is a mean trace distance between the sampled state $\rho$ and the mean sampled state $\langle \rho \rangle$. 

\end{document}